**OncoScore: a novel, Internet-based tool to assess the oncogenic potential of genes**

Authors:

Piazza Rocco[1], Ramazzotti Daniele[2], Spinelli Roberta[1], Pirola Alessandra[3], De Sano Luca[4], Ferrari Pierangelo[3], Magistroni Vera[1], Cordani Nicoletta[1], Sharma Nitesh[5], Gambacorti-Passerini Carlo[1]

1) University of Milano-Bicocca, Dept. of Medicine and Surgery, Monza, 20900, Italy

2) Stanford University, Dept. of Pathology, California 94305, USA

3) GalSeq s.r.l., viale Italia 46, Monza, 20900, Italy

4) University of Milano-Bicocca, Dept. of Informatics, 20125, Milano

5) University of New Mexico, Department of Pediatrics, Albuquerque

E-mail:

Piazza Rocco: rocco.piazza@unimib.it

Ramazzotti Daniele: daniele.ramazzotti@stanford.edu

Spinelli Roberta: spinelli.roberta@unimib.it

Pirola Alessandra: alessandra.pirola@galseq.com

De Sano Luca: luca.desano@gmail.com

Ferrari Pierangelo: pierofer@hotmail.it

Magistroni Vera: vera.magistroni@unimib.it

Cordani Nicoletta: nicoletta.cordani@unimib.it

Sharma Nitesh: NDSharma@salud.unm.edu

Gambacorti-Passerini Carlo: carlo.gambacorti@unimib.it

Corresponding author: Rocco Piazza

The complicated, evolving landscape of cancer mutations poses a formidable challenge to identify cancer genes among the large lists of mutations typically generated in NGS experiments. The ability to prioritize these variants is therefore of paramount importance. To address this issue we developed OncoScore, a text-mining tool that ranks genes according to their association with cancer, based on available biomedical literature. Receiver operating characteristic curve and the area under the curve (AUC) metrics on manually curated datasets confirmed the excellent discriminating capability of OncoScore (OncoScore cut-off threshold = 21.09; AUC = 90.3%, 95% CI: 88.1-92.5%), indicating that OncoScore provides useful results in cases where an efficient prioritization of cancer-associated genes is needed.

**Introduction**

The huge amount of data emerging from NGS projects is bringing a revolution in molecular medicine, leading to the discovery of a large number of new somatic alterations that are associated with the onset and/or progression of cancer. However, researchers are facing a formidable challenge in prioritizing cancer genes among the variants generated by NGS experiments. Despite the development of a significant number of tools devoted to cancer driver prediction, limited effort has been dedicated to tools able to generate a gene-centered Oncogenic Score based on the evidence already available in the scientific literature. To overcome these limitations, we propose here OncoScore, a bioinformatics text-mining tool capable of automatically scanning the biomedical literature by means of dynamically updatable web queries and measuring gene-specific cancer association in terms of gene citations. The output of this analysis is a score representing the strength of the association of any gene symbol to cancer, based on the literature available at the time of the analysis. OncoScore is distributed as a R Bioconductor package (https://bioconductor.org/packages/release/bioc/html/OncoScore.html) in order to allow full customization of the algorithm and easy integration in existing NGS pipelines, and as a web tool for easy access by researchers with limited or no experience in bioinformatics (http://www.galseq.com/oncoscore.html).

**Results**

We analyzed the performance of OncoScore on the Cancer Genes Census (CGC; Supplementary Tab. 1), a collection of regularly updated and manually annotated genes accepted as causally implicated in oncogenesis [1]. To assess the ability of OncoScore to discriminate between cancer and non-cancer genes we generated the OncoScore estimation for the whole CGC dataset and for a manually curated list of genes not associated with cancer (named nCan; Supplementary Tab. 2; see Methods section for further details). Genes with a total citation count < 10 publications were filtered out, therefore from a total of 507 CGC and 302 nCan, 472 (93.1%) and 266 (88.1%) genes were further processed.

**OncoScore performance**

The distribution of OncoScore values differed significantly between the two groups (mean: 48.8 and 14.8 for CGC and nCan, respectively; $p$-value = $2.2e^{-16}$; Fig. 1a-b). The receiver operating characteristic (ROC) curve and the area under the curve (AUC) metrics (Fig. 2a-b) confirmed the excellent capability of OncoScore in discriminating the true positive from the true negative cancer genes at different cut-off values (OncoScore cut-off threshold = 21.09; AUC1 = 90.3%, 95% CI: 88.1-92.5; see Methods section for further details). The same analysis performed on the entire list of known human genes (Supplementary Tab. 3) using an identical cut-off (21.09) identified a total of 5945 cancer-related genes, corresponding to 35% of the total (Suppl. Fig. 1).

**Comparative analysis**

The results obtained by OncoScore were then compared with those of Gene Ranker (http://cbio.mskcc.org/tcga-generanker/) [2], a web tool extensively used to estimate the oncogenicity of putative cancer genes. Gene Ranker assigns a cancer score to each gene according to a complex set of fixed properties (details can be found at: http://cbio.mskcc.org/tcga-generanker/sources.jsp), where the score associated with each gene is based on the confidence of the data used to generate the prediction, therefore, from a functional point of view OncoScore and Gene Ranker are very similar. To compare the performance of the two approaches we generated the Gene Ranker ROC curve and we calculated the AUC metric (AUC2 = 85.98%, 95% CI: 0.8337-0.886). Gene Ranker was able to score 452/507 (89.2%) CGC and 294/302 (97.4%) nCan genes for a total of 746 out of 809 genes. Globally, OncoScore showed greater AUC (AUC1 > AUC2) than Gene Ranker (see Methods section for further details). Notably, a significant number of well-known oncogenes/oncosuppressors, such as *ASXL1*, *SETBP1*, *TET2*, *SF3B1* and *SRSF1* received an extremely low Gene Ranker score (< 0.250) or were not classified at all, which is probably explained by the use of time-fixed scoring criteria (Fig. 1c-d).

**Chronic myeloid leukemia**

Subsequently, we tested OncoScore on real NGS data obtained from patients affected by Chronic Myeloid Leukemia (CML). CML originates as an indolent disease with limited aggressiveness (chronic phase, CP). However, if left untreated, it invariably progresses to a final, aggressive phase known as blast crisis (BC). The CML-CP is causally associated with the Breakpoint Cluster Region (BCR) and Abelson (ABL) gene fusion (BCR-ABL) [3,4]. The presence of the BCR-ABL fusion is necessary and sufficient to generate the leukemic phenotype, therefore a very limited number of oncogenic variants besides BCR-ABL are expected to be present in CP at diagnosis, at least in the predominant clone. In addition, since CML is a disease affecting mostly old people, CML-CP cells are expected to carry a significant number of passenger variants [5]. Conversely, evolution towards BC is caused by a set of only partially characterized oncogenic variants responsible for the impairment of the myeloid differentiation machinery [6]. BC evolution usually occurs within 2-3 years after the onset of the disease, hence the accumulation of passenger variants during the evolution from CP to BC is limited. Therefore, in CML the number of somatic mutations functionally associated to cancer is expected to be higher in BC than in CP.

To test OncoScore, we generated whole-exome sequencing data on 33 Chronic Myeloid Leukemia patients, 23 in Chronic Phase and 10 in Blast Crisis. In order to selectively identify somatic mutations, CP was filtered against the germline exome data, while BC was filtered against the corresponding CP exome (see Methods section for further details). A total of 107 and 34 nonsynonymous somatic variants were identified in CP and BC, respectively (Supplementary Tab. 4). All the variants were validated by Sanger sequencing in case and control samples. The OncoScore estimation was subsequently performed for all the 141 Sanger-validated somatic variants (Supplementary Tab. 4): the score of the BC associated variations was higher than that of CP (mean OncoScore = 35.6 ± 4.91 SEM and 19.2 ± 2.07 SEM for BC and CP, respectively; p=0.0007; Fig. 3). Manual inspection of the top 10 BC OncoScore genes (Supplementary Tab. 5) highlighted the presence of at least 5 genes (*ABL1*, *NRAS*, *ASXL1*, *RUNX1*, *IKZF1*) that are functionally associated with CML progression [7,8], suggesting that OncoScore was able to correctly prioritize biologically relevant CML cancer genes.

**OncoScore analysis of large structural variations**

Large deletions and amplifications are among the most frequent structural variations occurring during carcinogenesis. Unfortunately, the identification of the underlying oncogenes/oncosuppressors is very challenging. The reason is that the size of the structural variants is often in the range of the megabases, therefore involving a large number of genes. To allow a fast and efficient prioritization of driver genes involved in large copy-number abnormalities, OncoScore is able to process chromosomal regions as query input and to output a list of all the genes present in the input region, together with the associated OncoScore. This allows to quickly verify if some of the genes involved in the structural variation can play an important role in cancer onset/evolution. To test this feature we analyzed three regions (Supplementary Table 6), 8p21, 10q23 and 12p13, known to be recurrently deleted in prostate cancer [9,10]. Manual analysis of the literature indicates that *NKX3.1*, *PTEN* and *CDKN1B* are important oncosuppressors that are lost upon deletion of the 3 loci [9-13] and are known to play a driver role in prostate cancer evolution.

For the sake of keeping the analysis focused on the genes with the highest driver potential, we initially removed all the events with an OncoScore of 0 and then we selected those with an OncoScore estimation above the 90$^{th}$ percentile. Overall, OncoScore analysis of the 8p21 region revealed the presence of a total of 106 events. Of them 74 had an OncoScore > 0 and 47 an OncoScore ≥ 21. A total of 8 cancer genes with an OncoScore above the 90$^{th}$ percentile was identified (*TNFRSF10B*, *TNFRSF10C*, *TNFRSF10A*, *TNFRSF10D*, *LZTS1*, *NKX3-1*, *BNIP3L* and *RHOBTB2*; OncoScore range: 71.4-86.3). The same analysis done on the 10q23 deletion identified a total of 124 events, 69 with an OncoScore > 0, 39 with OncoScore > 21 and 7 above the 90$^{th}$ percentile (*PTEN*, *KLLN*, *SNCG*, *TNKS2*, *CEP55*, *PLCE1 and HELLS*; OncoScore range: 50.6-74.1). A total of 256 events was identified in the 12p13 locus. Of them 141 had an OncoScore > 0 and 85 an OncoScore ≥ 21. A total of 14 cancer genes with an OncoScore above the 90$^{th}$ percentile was identified (*ETV6*, *TNFRSF1A*, *ING4*, *CDKN1B*, *MIR200C*, *LTBR, FOXM1, GPRC5A, KDM5A, TIGAR, PRB2, CCND2, RAD51AP1 and KLRK1*; OncoScore range: 59.1-86.7).

Notably, the 3 known driver genes, *NKX3.1*, *PTEN* and *CDKN1B*, were all present in the OncoScore list, ranking 6$^{th}$, 1$^{st}$ and 4$^{th}$, respectively. Moreover, several other genes present in the OncoScore lists, such as *KLLN*, *LZTS1*, *BNIP3L*, *ING4* as well as *MIR200C* microRNA were recently implicated in the development of prostate cancer [14-18]. Taken globally these data indicate that OncoScore is a useful prioritization tool for the annotation of copy-number abnormalities in human cancer.

**Time-series plot**

Given the strong impulse generated by NGS to cancer research, it is not uncommon that a gene previously considered 'non-cancer' subsequently turns-out to be a driver. The OncoScore of a newly discovered cancer gene will start increasing over time, as its oncogenic role is confirmed by subsequent studies. This process, however, requires a significant amount of time, causing a potential delay between the identification of the oncogene and the acquisition of a 'driver' OncoScore annotation. To facilitate the identification of recently discovered cancer genes we implemented a time-series function (*perform.time.series.query*) which allows plotting the OncoScore through a user-defined time-window. To test this function, we generated time-series queries spanning from 1975 to 2016 (Fig. 4) and involving two different dataset: 1) a set of manually defined cancer genes (*TP53*, *KRAS*, *NRAS*, *HRAS*, *ASXL1*, *IDH1*, *IDH2*, *TET2*, *SETBP1*) as well as the housekeeping genes *GAPDH* and *GUSB*; 2) a set of 10 genes randomly extracted from the CGC (*ARID1A*, *HMGA2*, *KIF5B*, *NUP214* and *RBM15*) and nCan (*ALMS1*, *DCAF17*, *GPD1L, WFS1* and *RBM10*) lists. In dataset 1, at the final time point (2016), all the cancer genes under analysis scored > 60, while the two housekeeping reached a plateau at ~20. The dynamics of the OncoScore pattern revealed the presence of 2 cancer gene clusters: the first one comprising oncogenes/oncosuppressors identified in the 1985-94 decade (*TP53* and the RAS family), right after the development of the PCR by Kary Mullis [19] and just a few years after the invention of the 'Sanger' sequencing technique by Frederick Sanger [20]; the second one occurring right after the NGS breakthrough and comprising *ASXL1*, *IDH1/2*, *TET2* and *SETBP1*. In particular, the behavior of SETBP1 curve is interesting, because it reflects the complex story of SETBP1 discovery. SETBP1 was initially identified as an oncogene (NUP98-SETBP1 fusion) in pediatric acute T-cell lymphoblastic leukemia by Panagopoulos and colleagues [21], which explains the first, sharp increase in SETBP1 OncoScore back in 2007. Subsequently, in January 2010, Cristobal and colleagues [22] demonstrated SETBP1 overexpression as a novel leukemogenic mechanism in acute myeloid leukemia. Their finding is represented as a second peak in the SETBP1 time-series. Their finding was shortly followed by a seminal publication by Hoischen and colleagues [23] where the authors demonstrated that de novo, germline SETBP1 mutations were responsible for the onset of the Schinzel-Giedion syndrome (SGS), a severe disorder characterized by severe mental retardation, distinctive facial features and multiple congenital malformations. Given that this finding doesn't directly associate SETBP1 with cancer, this led to a decrease in the overall SETBP1 score, because in the period between 2010 and 2013 a number of papers confirming the link between SETBP1 and SGS (and therefore counting as negative for the OncoScore) appeared in the literature. Despite this decrease however, SETBP1 score never fell below the OncoScore cut-off threshold, therefore remaining in the cancer-associated genes group. Finally, in 2013 we [24] and others [25,26] demonstrated the occurrence of somatic, oncogenic SETBP1 point mutations in several types of cancer, which caused a new increase in the overall SETBP1 OncoScore. In the second dataset all the CGC genes (5/507) were classified as oncogenes by OncoScore and 4 (4/302) out of 5 nCan genes (*ALMS1*, *DCAF17*, *GPD1L* and *WFS1*) were classified as 'non-cancer' at the final time point, as expected. The remaining nCan gene (1/302), *RBM10*, showed an interesting behavior, as its OncoScore remained close to 0 until 2010-2011, where it arose abruptly to over 40. Manual analysis of the literature showed a very recent association of this gene with cancer [27-29], which highlights the usefulness of the OncoScore analysis in order to identify recently discovered cancer-associated genes.

**Discussion**

The progress of genomic-scale sequencing allows the analysis of cancer genomes at an unprecedented resolution, significantly increasing our ability to study the mechanisms underlying cancer onset and progression. Many issues remain however to be solved in order to transform an apparently meaningless list of mutated genes into useful information. One of these issues is represented by the ability to differentiate genes with oncogenic potential involved in the early stages of neoplastic transformation from passenger mutations. Both are present in the majority of tumor cells being analyzed; therefore differences in mutation frequencies cannot be utilized here. Despite dramatic progresses in sequencing technologies, a systematic, painfully slow manual analysis of cancer literature remains the main approach by which biomedical researchers try to assess the oncogenicity of genes identified in cancer samples and to prioritize them. In this study, we proposed a new method aimed at quickly identifying cancer-associated genes through automated PubMed queries and dedicated text-mining algorithms. To perform PubMed queries we took advantage of the public API to the NCBI Entrez system: the NCBI E-utilities (http://www.ncbi.nlm.nih.gov/books/NBK25500/).

By comparing OncoScore with Gene Ranker we demonstrated that our tool has better discriminating capabilities than the latter.

At present, possible issues affecting OncoScore are:

1) Analysis of genes with a very limited number of bibliographic references: limited data-mining information represents a significant issue, because it causes a significant loss of classification power. Therefore, under standard OncoScore settings, genes with less than 10 references are not considered. This problem is mitigated by the accumulation of new data causing the number of genes with limited or no bibliographic references to shrink over time.

2) A query error can occur because of the presence of 'ambiguous' gene names. Indeed, HGNC names would be preferable over gene symbols given their uniqueness. Unfortunately PubMed expects gene symbols in standard queries and doesn't allow the use of symbol tags. Given the process of gene symbol definition, a minimal fraction of ambiguous cases is therefore unavoidable when using symbols as the primary input. To limit this potential issue however, OncoScore will allow the end-user to: 1) directly input GeneID numbers instead of symbols; 2) automatically check if a gene symbol is ambiguous and to prompt an alert in order to allow the end-user to disambiguate it.

3) Another problem may exist for genes falling in the uncertainty region close to the optimal cut-off value (21.09). For instance, housekeeping genes such as *GAPDH* (OncoScore: 23.29) and *GUSB* (OncoScore: 20.53) fall within this group. These genes are often cited in cancer biological studies as reference genes used for normalization, which suggests that categorization of genes with an OncoScore close to the threshold should be interpreted with caution.

**Conclusions**

In this work we presented OncoScore, a bioinformatics data mining tool dedicated to the assessment of the oncogenic potential of genes. OncoScore performance improved the state-of-the-art on manually curated datasets, therefore it could be useful in all cases where an efficient differentiation between potential drivers and passenger genes is needed. The strategy used to implement OncoScore allows the model to be continuously up to date, easy to use and executable in real-time. OncoScore analysis on real NGS variant data as well as on chromosomal regions shows the utility of this tool in the crucial task of cancer gene prioritization. The identification of genes causally involved in cancers pathogenicity will be greatly helped by OncoScore, although the final assignment of a driver gene to an individual cancer will require further information such as the histological type and the expression level of the mutated gene as well as information on the functional effect of the mutation.

**Methods**

**Gene Datasets**

The CGC dataset was collected from the Cancer Gene Census (CGC; version: November 2013; http://www.sanger.ac.uk/science/data/cancer-gene-census) database, a set of genes causally implicated in cancer. The CGC (November 2013) accounts for 507 cancer-associated genes. The not-CGC (nCan) were collected from the GeneCards database and represent a set of genes not previously involved in the onset and in the progression of neoplastic diseases although they may be involved in human genetic and syndromic diseases. nCan were manually selected from a list of genes downloaded from Gene Cards (Release 3.10) according to the Gene Cards description.

**Data Sources**

Gene citations, used as control, and cancer related genes citations from biomedical literature were retrieved from PubMed repository. PubMed is a freely accessible web search engine developed and maintained by the National Center for Biotechnology Information (NCBI) at the National Institutes of Health (NIH) and it is updated daily. PubMed accounts for more than 24 million citations of biomedical literature from MEDLINE (http://www.ncbi.nlm.nih.gov/pubmed), from life science journals and from online books. The citations were retrieved from PubMed by using a set of R scripts developed according to the Entrez Programming Utilities (E-utilities) [30]. The E-utilities are an interface to the Entrez system, which includes many databases covering a variety of biomedical data such as the biomedical literature.

**Text Mining: Keywords search terms and syntax queries**

In order to retrieve the citation counts, dedicated text-mining R scripts were developed based on optimized keywords. The genes citations numbers were searched by the gene symbols while the cancer genes citations were searched by the gene symbols associated with the following keywords search terms "(*lymphoma or tumor or neoplasm or malignancy or cancer or leukemia*)". The aim of such a set of keywords is to give a comprehensive definition of the topic in order to retrieve all the articles concerning cancer research. In both cases, the search terms were processed according to the "Automatic Term Mapping", a process used by PubMed to find a match to unqualified terms that are entered into the search field. Untagged terms are matched (in this order) against subjects using the MeSH (Medical Subject Headings) translation table, journals using the Journals translation table, and authors and investigators, using the the Full Author translation table, Author index, Full Investigator translation table and Investigator index. If a match is found in any translation table, the mapping stops. When subject or journal matches are found, the query and individual terms are also searched in All Fields. If no match is found in any tables, terms are searched in All Fields and ANDed together (see https://www.nlm.nih.gov/bsd/disted/pubmedtutorial/020_040.html).

**The OncoScore Analysis**

The OncoScore analysis consists of a main module aimed at assessing the oncogenic potential of genes. It makes use of the citations retrieved from the literature to calculate the score and to classify genes with the aim of discriminating cancer from not-cancer putative genes.

**OncoScore Definition**

OncoScore is a measure (i.e. a score) of association of genes with cancer based on gene citation frequency from biomedical literature, here labelled $CitGene$, and gene citation frequency linked to cancer, labelled $CitGene\ AND\ Cancer$. The OncoScore is defined to be directly proportional to the percentage of gene cancer citations, labelled $PercCit$, as (1) $PercCit = \frac{CitGene\ AND\ Cancer}{CitGene} \cdot 100$. From this, OncoScore is defined as (2) $Score = PercCit - \alpha \cdot PercCit$ where $\alpha > 0$ and it is weighted for the gene citation value (3) $\alpha = \frac{1}{Log_2(CitGene)}$ where $CitGene > 1$. So, the OncoScore is computed as follow (4):

$$OncoScore = \frac{CitGene\ AND\ Cancer}{CitGene} \cdot 100 - \frac{1}{Log_2(CitGene)} \cdot \frac{CitGene\ AND\ Cancer}{CitGene} \cdot 100.$$

The OncoScore is, hence, based on the number of PubMed articles in which both the gene symbol and the cancer keywords appear in the same article, weighted for the logarithmic gene citation number as defined above.

The OncoScore discriminative power was assessed on CGC (Supplementary Tab. 1) and nCan (Supplementary Tab. 2) datasets. The OncoScore distribution among CGC and aCan genes was assessed by Mann-Whitney-Wilcoxon Test and visualized by boxplot (Fig. 1a-b).

**OncoScore Classifier**

In order to identify a general discriminating model based on the current information available from the literature, we tested different automated classifiers trained on the panels of CGC and nCan genes. Specifically, we adopted classifiers built by Linear Discriminant Analysis (LDA) and Logit analysis (LOGIT) to select an optimal model in terms of classification accuracy that we dubbed OncoScore classifier model. The CGC and not-CGC genes were used to model the probabilities of binary outcomes as a function of the OncoScore by using linear discriminant and logistic functions in the presence of two classes of categorical data, i.e. oncogene vs not-oncogene. The best prediction accuracy, 84.96% ($ACC_1$), was achieved at the optimal OncoScore cut-off equal to 21.09, obtained both by Youden's and top left corner methods [31] (sensitivity 89.32%; specificity 77.78%; Fig. 2a,b).

**OncoScore Classifier assessment**

The CGC and nCan genes were analyzed to compare OncoScore results with Gene Ranker. The best Gene Ranker discriminating threshold for CGC and nCan was calculated considering the best between the Youden's and top left threshold by maximizing the classification accuracy. Then, cancer associated and non-associated genes were discriminated according to the best Gene Ranker threshold. The accuracy, sensitivity and sensibility were inferred at the optimal threshold and they were compared to the best OncoScore discriminating model.

**Time Series OncoScore Profile**
The Time Series OncoScore analysis computes scores over time by making use of the citations from the literature at different time points. The dynamic score aims to model the association of genes to cancer over time, following the evolution of the literature in order to assess the trend to which the OncoScore tended to move. The citations over the time series are retrieved by automatically performing data mining queries at various time points.

**Patients Samples**
A total of 23 CP-CML and 10 BC patients were analyzed. Mononuclear cells were collected at diagnosis from bone marrow using Ficoll density gradient centrifugation (tumor sample, CP-CML) and peripheral blood or bone marrow mononuclear cells were collected after achievement of Complete Cytogenetic Remission (control sample). For BC samples, mononuclear cells were collected at the time of progression using Ficoll density gradient centrifugation (BC-CML). For those patients who never achieved CCyR, normal lymphocytes were expanded in vitro with IL-2 and phytohemagglutinin, as previously published [24]. Informed consent was obtained from all subjects. Experimental protocols were approved by the University of Milano-Bicocca institutional committee. All methods were performed in accordance with relevant guidelines and regulations.

**Whole Exome Sequencing**
Genomic DNA (gDNA) was extracted from purified cells with PureLink Genomic DNA kit (Life Technologies-Thermo Fisher). Two micrograms of gDNA were sheared to a size of 200-300bp using a Bioruptor NextGen Sonicator (Diagenode) and processed according to the standard Illumina protocol. The Illumina TruSeq Exome Enrichment kit was used to enrich the genomic libraries for the exonic regions. Matched exome libraries of tumor and non-tumor DNA were prepared by Illumina TruSeq DNA Sample Preparation Kit and Illumina TruSeq exome enrichment kit by using a standard protocol. All the captured DNA libraries were sequenced on an Illumina HiSeq 2500 or a Genome Analyser IIx in paired-end mode to yield 76bp sequencing reads.

**Base Calling and Sequence alignment**
Base calling and quality control were executed on the Illumina RTA sequence analysis pipeline (Illumina). Paired Fastq files were aligned to the human reference genome (GRCh37/hg19) with the Burrows–Wheeler-based BWA alignment tool [32] using the BWA-MEM algorithm. Duplicated reads were excluded from the analysis using Samblaster. Quality of the aligned reads, somatic variant calling and copy number analysis were performed using CEQer2, an in-house evolution of CEQer tool [33] as previously described [24,34]. Splicing variants were analyzed using SpliceFinder [35]. Variants were annotated using dbSNP146. All the filtered variants, exported as vcf files, were annotated using Annovar [36] and manually inspected.
Variants present in less than 35% of the tumor reads were filtered-out, in order to specifically isolate the variants present in the major clone. All the candidate somatic mutations identified by WES were then re-confirmed by Sanger sequencing.

**Validation of mutations by Sanger sequencing**
Variant DNA was amplified by PCR (FastStart High Fidelity PCR system, Roche Applied Science, Germany) and the amplicons of tumor and normal DNA were processed for Sanger sequencing at GATC (GATC Biotech AG, Cologne, Germany). The presence of each substitution was confirmed by using Chromas platform (Technelysium).

Acknowledgements

We kindly acknowledge the contribution of Michela Viltadi for technical help. This work was supported by Associazione Italiana Ricerca sul Cancro 2013 (IG-14249 to C.G.P.), by Associazione Italiana Ricerca sul Cancro 2015 (IG-17727 to R.P.), by Fondazione Berlucchi (2014) and by European Union's Horizon 2020 Marie Skłodowska-Curie Innovative Training Networks (ITN-ETN) under grant agreement No.: 675712CGP; CGP is a member of the European Research Initiative for ALK-Related Malignancies (www.erialcl.net).

Authors' contribution

Rocco Piazza conceived research, analyzed data, and wrote the manuscript

Daniele Ramazzotti wrote the OncoScore code

Roberta Spinelli wrote the OncoScore code

Alessandra Pirola generated next generation sequencing data

Luca De Sano wrote the OncoScore code

Pierangelo Ferrari developed the OncoScore Web Tool

Vera Magistroni validated NGS data by Sanger sequencing

Nicoletta Cordani analyzed data and critically revised the manuscript

Nitesh Sharma validated NGS data by Sanger sequencing

Carlo Gambacorti-Passerini conceived research and critically revised data

Competing interests

The authors declare that they have no competing interests.

Fig. 1. OncoScore distribution of 'Cancer' and 'Non-Cancer' gene sets. BoxPlot (a) and frequency histogram (b) of the OncoScore distributions for non-cancer and cancer genes. (a) Each box plot is drawn between the lower and upper quartiles of the distributions with bold black line showing the median value. The OncoScore distributions of 'Cancer' and 'Non-Cancer' genes are significantly different (Mann-Whitney-Wilcoxon Test: p-value = 2.2e-16). b) OncoScore frequency distribution plotted by equispaced breaks. c) OncoScore and d) Gene Ranker ranking plot of a mixed panel comprising 'Cancer' (*) and 'Non-Cancer' genes. The horizontal red lines identify the best cut-off classifier threshold models.

Fig. 2. a) OncoScore prediction accuracy. ROC curve depicting the relationship between true positive rate (Sensitivity) and true negative rate (Specificity) and AUC metric on CGC and nCan genes. b) OncoScore density score distribution of true positives and true negatives. The blue line represents the CGC and the grey one the nCan genes. The dashed red line shows the optimal Youden's cut-off threshold.

Fig. 3. Boxplot reporting the OncoScore values of all the genes carrying somatic mutations in chronic phase or blast crisis chronic myeloid leukemia samples. P-value = 0.0007 (Two-tailed Mann-Whitney test).

Fig. 4. Time-series OncoScore plot spanning from 1975 to 2016. A) Time-series plot involving a set of manually defined cancer (*TP53*, *KRAS*, *NRAS*, *HRAS*, *ASXL1*, *IDH1*, *IDH2*, *TET2* and *SETBP1*) and housekeeping genes (*GAPDH* and *GUSB*). The grey boxes highlight two major scientific breakthroughs occurred during this time span. B) Time-series plot of 10 genes randomly selected from the CGC (*ARID1A*, *HMGA2*, *KIF5B*, *NUP214*, *RBM15;* dashed lines) and nCan (*ALMS1*, *DCAF17*, *GPD1L*, *WFS1*, *RBM10*; continuous lines) dataset.

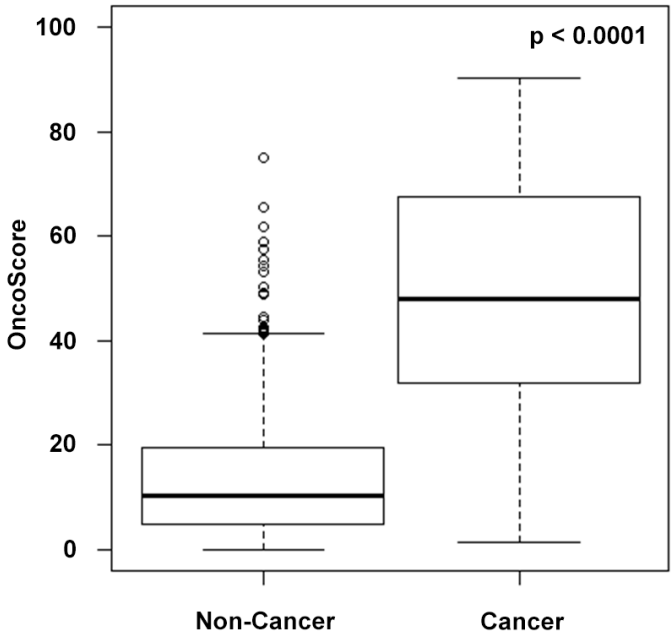
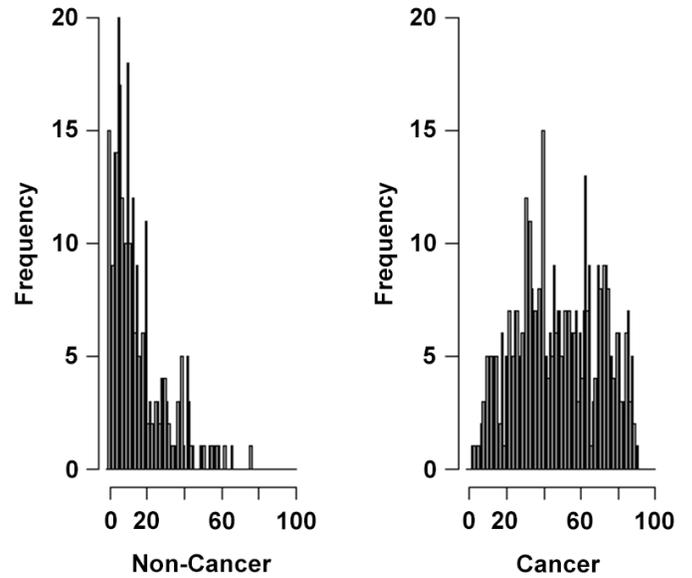
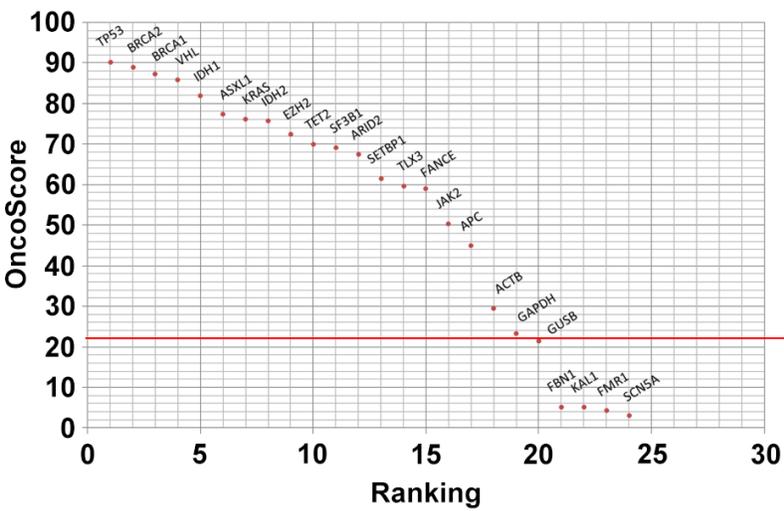
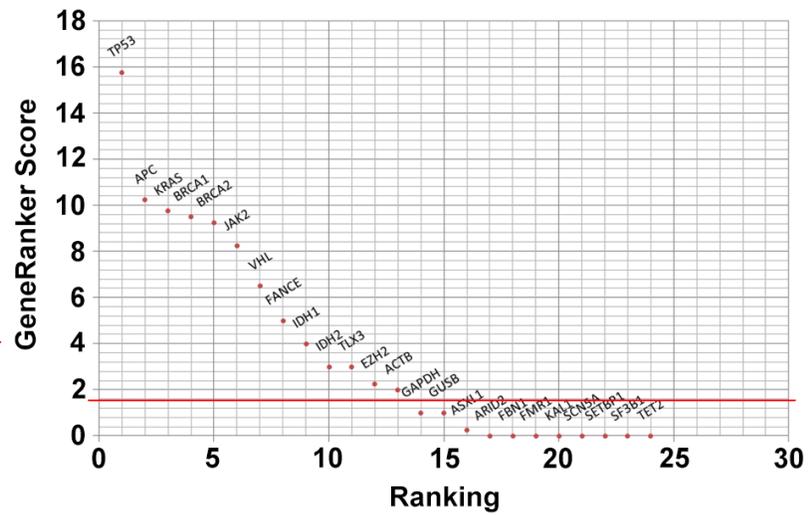

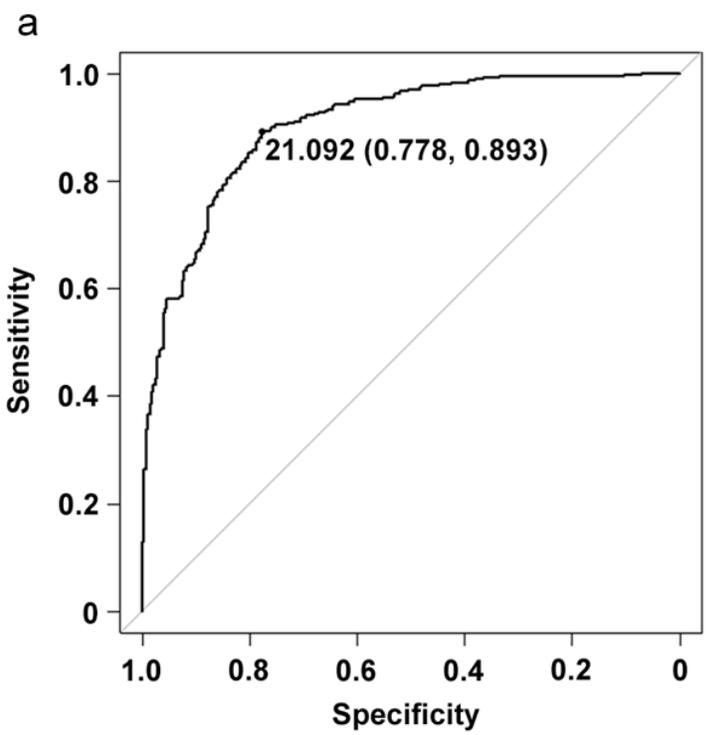 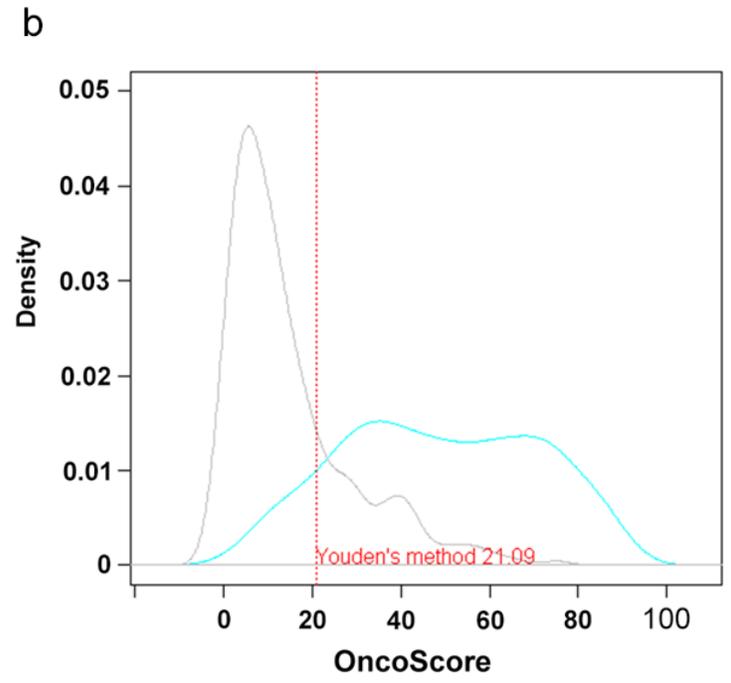

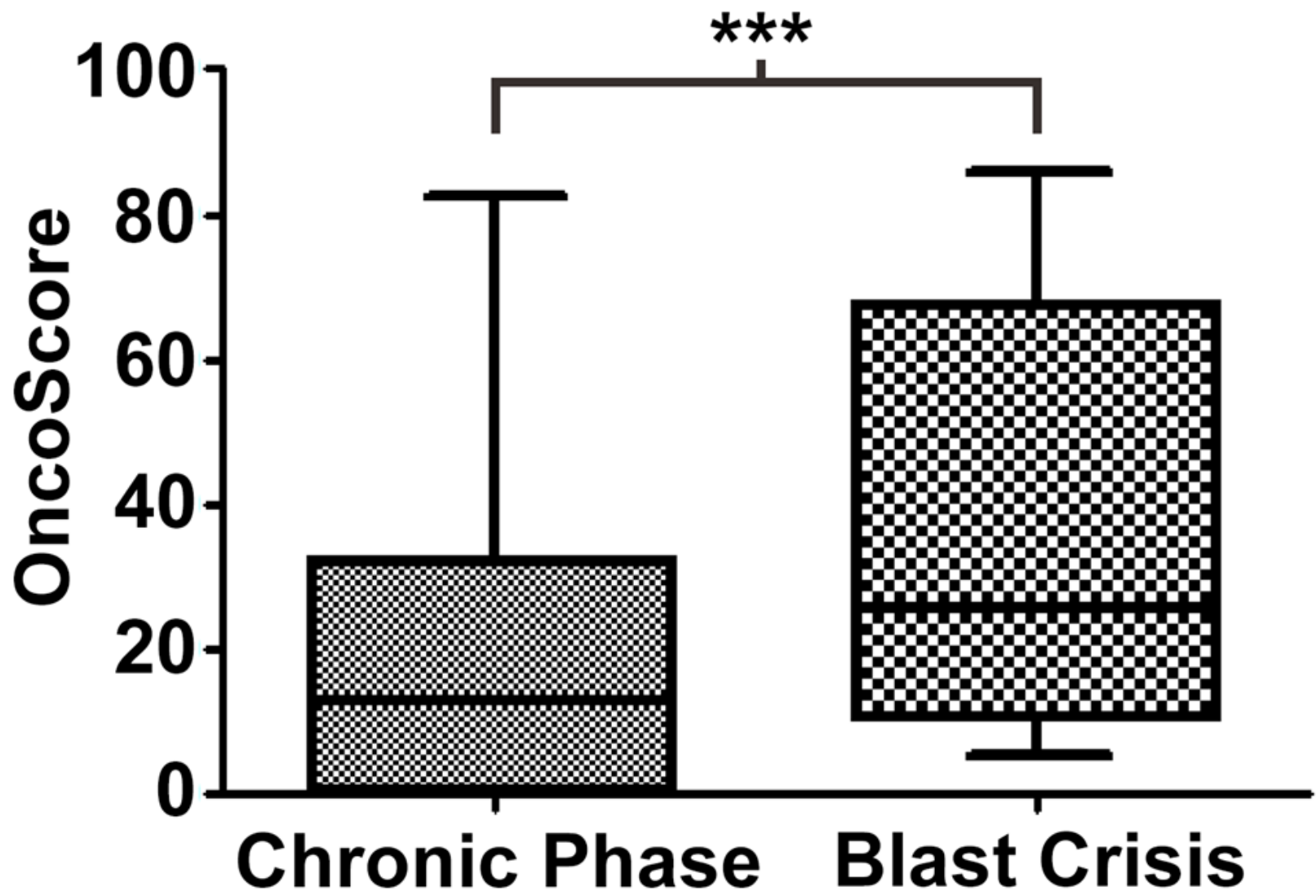

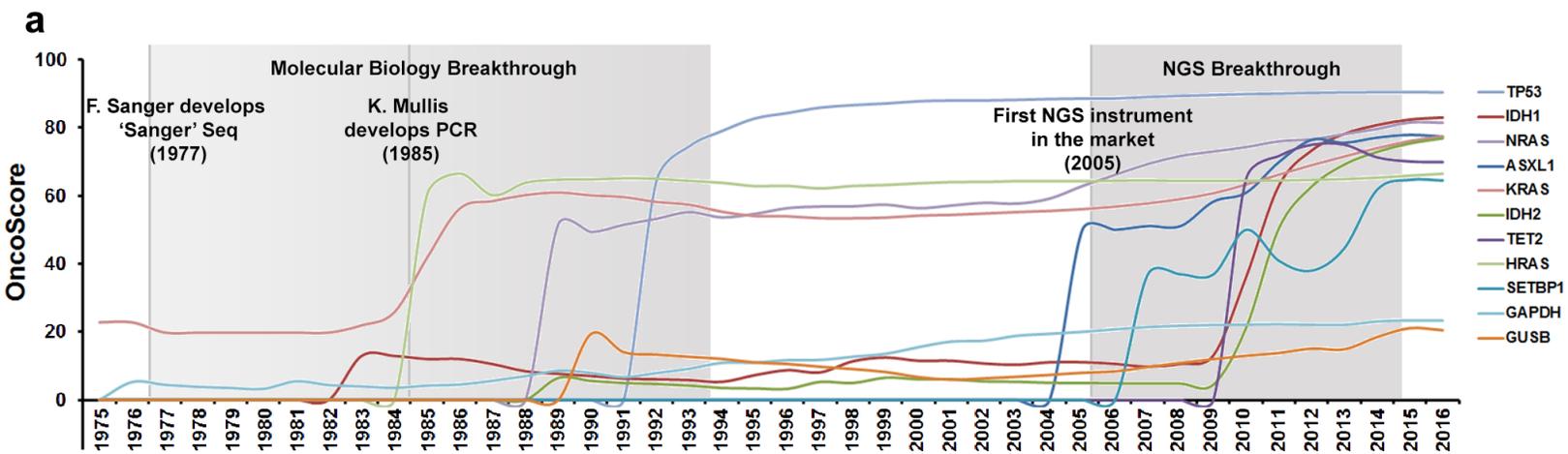
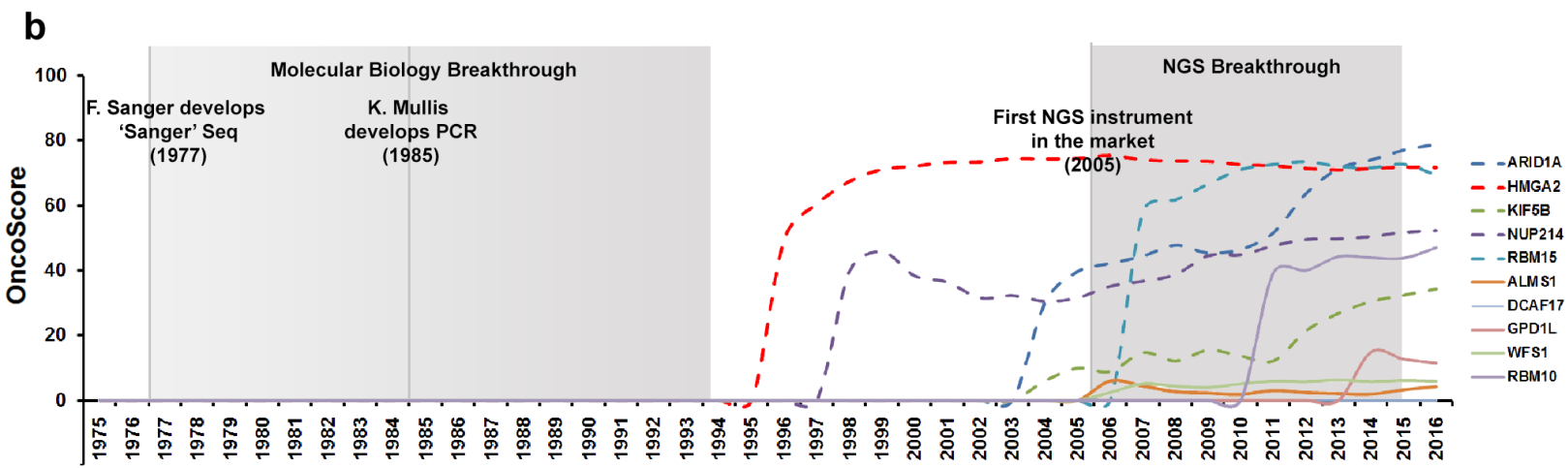